\documentclass[conference, 10pt]{IEEEtran}
\IEEEoverridecommandlockouts
\usepackage{cite}
\usepackage{amsmath,amssymb,amsfonts}
\usepackage{algorithmic}
\usepackage{graphicx}
\usepackage{textcomp}
\usepackage{xcolor}
\usepackage{setspace}
\usepackage{hyperref}
\usepackage{comment}
\usepackage{setspace}

\usepackage{fancyhdr}
\usepackage[norelsize, linesnumbered, ruled, lined, boxed, commentsnumbered, noend]{algorithm2e}

\begin{document}

\pagestyle{fancy}

\author{\IEEEauthorblockN{Suvarthi Sarkar, Abinash Kumar Ray, Aryabartta Sahu \textit{IEEE Senior Member}}
\IEEEauthorblockA{\textit{Dept. of CSE,  IIT Guwahati, Assam, India.} Emails: s.sarkar@iitg.ac.in, abinashkumarray@gmail.com, asahu@iitg.ac.in }
}

\title{Efficient Scheduling of Vehicular Tasks on Edge Systems with Green Energy and Battery Storage}

\setstretch{0.8}
\topmargin   -23mm
\textheight 249mm
\addtolength{\parindent}{-1.5mm}

\setlength\intextsep{1pt}%
\setlength\textfloatsep{2pt}

\addtolength{\abovecaptionskip}{-0.4cm}
\addtolength{\belowcaptionskip}{-0.5cm}
  
\maketitle

\begin{abstract}

The autonomous vehicle industry is rapidly expanding, requiring significant computational resources for tasks like perception and decision-making. Vehicular edge computing has emerged to meet this need, utilizing roadside computational units (roadside edge servers) to support autonomous vehicles. Aligning with the trend of green cloud computing, these roadside edge servers often get energy from solar power. Additionally, each roadside computational unit is equipped with a battery for storing solar power, ensuring continuous computational operation during periods of low solar energy availability.

In our research, we address the scheduling of computational tasks generated by autonomous vehicles to roadside units with power consumption proportional to the cube of the computational load of the server. Each computational task is associated with a revenue, dependent on its computational needs and deadline. Our objective is to maximize the total revenue of the system of roadside computational units.

We propose an offline heuristics approach based on predicted solar energy and incoming task patterns for different time slots. Additionally, we present heuristics for real-time adaptation to varying solar energy and task patterns from predicted values for different time slots. Our comparative analysis shows that our methods outperform state-of-the-art approaches upto 40\% for real-life datasets.

\end{abstract}


\setstretch{.9}

\section{Introduction}

The automotive sector is witnessing a significant surge in the development of autonomous vehicles (AVs), which are vehicles equipped with Autonomous Driving Assisted Systems (ADAS). The primary objective of AVs is to achieve an exceptionally high level of driving accuracy on roads \cite{Liu21}. Scientists believe that the key to attaining this goal lies in developing better cognitive models and utilizing more data to enhance environmental perception. These vehicles depend heavily on data gathered from their surroundings to operate without human drivers. While achieving fully autonomous vehicles remains an aspirational goal, substantial progress has been made compared to traditional manually operated vehicles.

Researchers have classified AVs into five levels, with level five representing vehicles capable of operating autonomously worldwide without human intervention. Currently, the AV industry has advanced to approximately level three. As AV technology continues to evolve, it is estimated that the data volume and computational requirements will increase significantly.

AVs progress from lower to higher automation levels, their data acquisition mechanisms undergo a significant transformation \cite{Soni22, Naceur16}. Initially relying on basic sensors for functions like adaptive cruise control, higher-level AVs integrate advanced technologies such as LiDAR for 3D mapping, radar for object detection, high-definition cameras for visual perception, and IMUs for precise localization \cite{Liu21, Ku21}. 
Additionally, AVs experience a surge in computational requirements for real-time decision-making in dynamic environments \cite{Ali20}. Advanced algorithms powered by AI and ML process vast sensor data, detect objects, predict hazards, plan routes, and adapt to various driving scenarios. However, managing the massive data generated by these sensors poses a computational challenge, necessitating high-quality computer chips that increase vehicle costs. This data is fundamental for AVs to navigate autonomously, highlighting the intricate balance between technological advancement and cost considerations in AV development \cite{Soni22}.

To address cost concerns and enhance affordability, AVs are leveraging infrastructure connectivity \cite{Dai22}. By connecting to infrastructure such as edge cloud computing platforms, AVs can offload computational demands, which we refer to as tasks, reducing the burden on their onboard computing resources. AVs execute these tasks in parallel: one in an onboard computation unit and another remotely in the infrastructure facility. These tasks encompass a range of functions like road safety services, automated overtaking, image processing tasks, boundary detection, and warning systems. They typically require a latency of only a few hundred milliseconds to ensure optimal performance \cite{Ku21, Ali20}. Edge cloud computing involves processing data closer to its source, which improves real-time processing capabilities and reduces latency. The ES is situated in a fixed location; it predominantly perceives approximately 80\% static information, with around 20\% being dynamic. The static information regarding the roadway can significantly enhance driving performance when communicated to the edge server \cite{cnn}. This approach enables AVs to handle complex tasks more efficiently without solely relying on onboard computing power. In this work, we refer to the infrastructure as Vehicular Edge System (VES).


This shift towards leveraging infrastructure for computational support addresses cost concerns and improves autonomous driving systems' overall efficiency and performance. Additionally, this integration of AVs with infrastructure paves the way for a more interconnected and intelligent transportation ecosystem, where vehicles, infrastructure, and data processing work seamlessly and collaboratively to ensure safe and efficient mobility for all.



The rapid advancement of technology has created a significant demand for computational power, driving growth in the cloud industry and increasing energy consumption. Green Cloud Computing (GCC) has emerged to address these challenges, reducing carbon footprints and supporting sustainable development while lowering operational costs  \cite{yamini}. Research indicates that nearly half of a data centre's power is consumed for cooling, prompting a shift towards colder climates, as exemplified by Google's relocation to Finland, which saved \$ 2.5 million \cite{powermodeldatacenter, cooldatacenter}. However, reliable power sources remain challenging in these areas, making solar energy a promising solution due to its universal accessibility.



In our work, we focus on leveraging solar energy as the primary source for performing computational tasks and increasing the revenue of the VES, as completion of tasks within the deadline is associated with revenue. However, we recognize that the availability of solar power can fluctuate, and it may not always be beneficial to execute tasks solely based on availability. To maximize the revenue of the VES, it needs to provide computational service to a maximum number of tasks and optimise the utilization of solar power while ensuring that we do not exceed the total solar power capacity available during a specific time slot. To achieve this goal, we develop a strategy that balances task execution with solar power availability. We aim to maintain a consistent overall utilization rate across all time stamps, which maximizes the utilization per unit of power consumed. This approach leads to increased revenue by efficiently utilizing available resources.

Another critical aspect of our strategy is using batteries to store surplus solar power. By storing excess energy in batteries during periods of high solar power generation, we can utilize this stored energy during periods of low solar power availability. This approach is useful as the power consumption increases non-linearly to the computational workload. So, using the power equally in mid to low quantities across all timeslots is beneficial.

Moreover, we consider the variability in task incoming rates. Since task arrivals can vary throughout the day, storing excess power in batteries is more beneficial than wasting it. 
Our approach not only maximizes the revenue of the infrastructure but also contributes to sustainability by reducing dependency on non-renewable energy sources. By efficiently managing solar power and battery storage, we demonstrate a practical and effective strategy for optimizing energy utilization in computational tasks.

Our contribution can be summarized as follows:

\begin{enumerate}
  \item We introduce an energy-aware scheduling approach designed for VES networks that rely on solar energy as their primary power source. Our work also incorporates using batteries to store surplus energy for later use during low solar energy availability.
  \item We proposed an optimal solution for the special case where solar power is used without a battery and tasks have equal computation demand and revenue.
  \item The proposed scheduling approach optimizes task scheduling and power allocation at different timeslots. It leverages knowledge of the incoming task set and solar energy profiles across different timeslots to maximise revenue.
  \item Our approach is adaptable to online scenarios where there may be slight deviations in task requirements and solar energy profiles compared to predicted data. We also conducted a comparative analysis against an offline approach to evaluate its performance, which yielded satisfactory results.
  \item To validate the effectiveness of our approach, we compared it with several state-of-the-art methods and assessed its performance across various parameters, demonstrating its superiority in maximised revenue generation.
\end{enumerate}

\section{Literature Review}

In edge computing systems (ECS), the equilibrium between computation offloading and energy efficiency is critical. Initial strides in this direction, such as the joint offloading and computing optimization by Wang et al. \cite{Wang2018}, provided foundational insights into managing computational tasks in wireless-powered ECS systems. Echoing these concerns, Chen et al. \cite{Chen2021} later revealed how task offloading, coupled with frequency scaling, can further enhance energy efficiency, demonstrating the potential of adaptable computational frequencies to minimize energy expenditure.

The relevance of such energy-efficient strategies extends to vehicular networks, where Li et al. \cite{Li2020} elucidated the significance of resource allocation in time-varying fading channels. Complementing this, Ku et al. \cite{Ku2020} presented real-time QoS optimization, crucial for maintaining service consistency, which, in turn, relies on optimized resource allocation and RSU selection, as illustrated by Li et al. \cite{Li2021}.

Beyond vehicular contexts, the concepts of energy-aware strategies in MEC have broad applications. HanLiang et al. \cite{HanLiang2020} delved into the energy dispatching strategies in data centres, emphasizing the importance of forecasting algorithms in managing energy resources effectively. This predictive approach to energy management resonates with the adaptive offloading framework proposed by Nan et al. \cite{Nan2017} for the cloud-of-things systems, highlighting a shift towards intelligent, anticipatory energy use.

Further works in this field, in general, explore advancements in energy harvesting \cite{Guo13, Nan2017}, the application of artificial intelligence for predictive resource management, and novel algorithms for efficient computation offloading. These studies build on and potentially synthesize the various research threads, from foundational theories to practical, scalable implementations, underscoring the dynamic interplay between energy efficiency and computational performance in the broader context of ECS and cloud computing.

In summary, the literature review demonstrates significant progress toward a more sustainable, efficient, and performance-oriented future in edge-cloud systems (ECS). Our work differentiates itself by considering the cubic power consumption model in relation to computational usage, a model that has not been addressed in previous studies. This model is highly relevant in modern data centres. Additionally, we conduct extensive experiments and analyses using both predicted and real-life data, addressing the dynamic nature of these environments—a factor often overlooked in existing literature.

\section{System Model and Problem
Statement}
\begin{figure}[tb!]
    \centering
    \includegraphics[scale = .35]{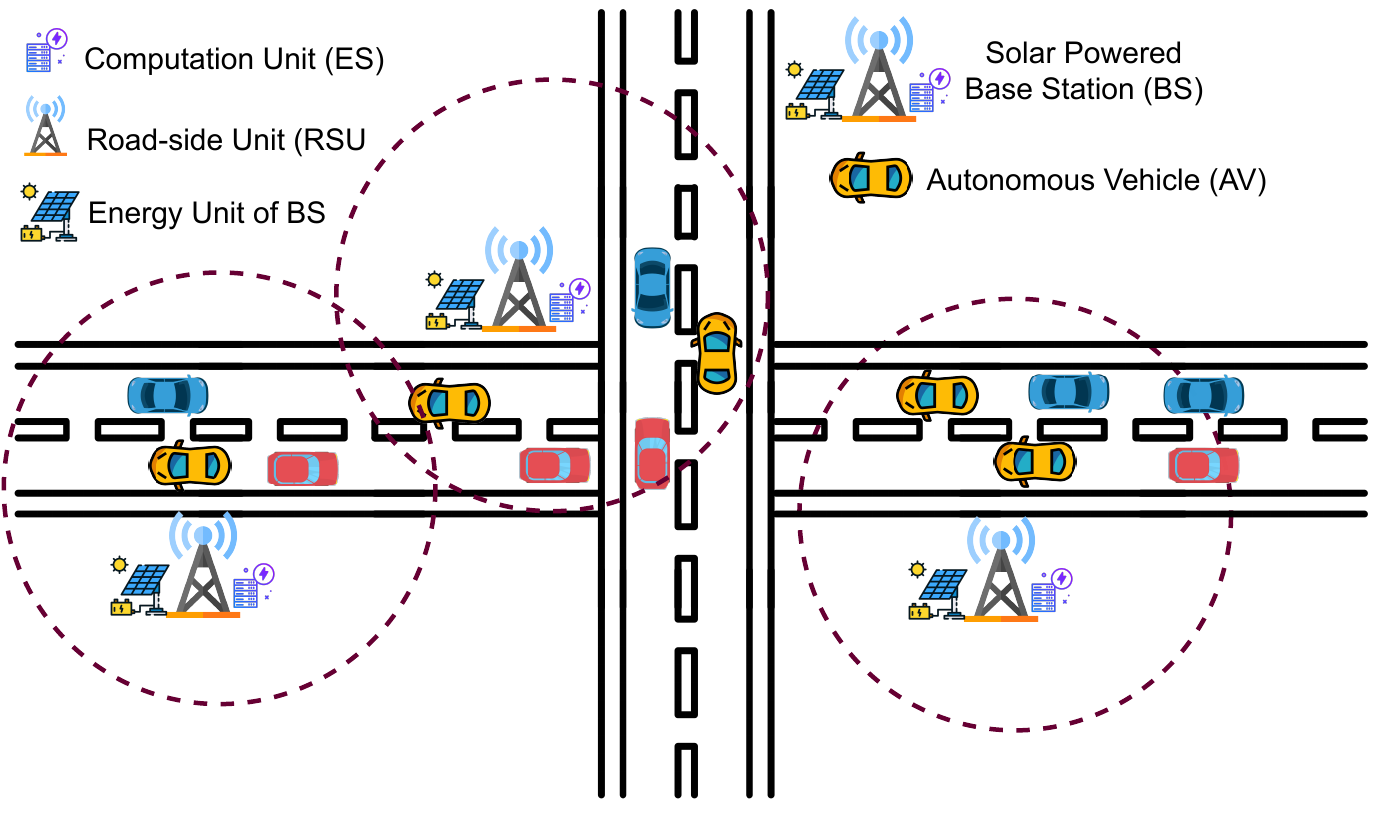}
    \caption{System Architecture}
    \label{fig:architecture}
\end{figure}
\subsection{System Model}
Our considered architectural model of Vehicular Edge
Systems (VES) as shown in \autoref{fig:architecture}. The considered VES consists of base stations (BSs) equipped
with edge servers (ESs). We consider there are $M$ BSs (or simply ES), $\boldsymbol{E} = \{E_1, E_2, \cdots, E_j, \cdots E_M\}$. We use index $j$ to uniquely identify each of the ESs. Every edge server possesses the capability to execute tasks concurrently across multiple cores. In this network, the vehicles are dynamic; we refer to them as autonomous vehicles (AVs). These AVs connect to the VES through these BSs and offload their computation requirements, and the ESs take care of the computation requirements in exchange for some monetary benefits. The BSs are strategically deployed to
ensure proximity to AVs, facilitating low-latency
communication. The BSs consist of Roadside Units (RSUs), which
act as communication gateways, enabling seamless interaction
between AVs and the VES. The ESs within BSs, to perform extensive computations, rely on a combination of attached solar panels and 
batteries to supply the required power. 

We consider each of the BSs is equipped with a battery and denote the batteries as $\boldsymbol{B} = \{B_1, B_2, \cdots, B_j, \cdots B_M\}$. The battery $B_j$ is associated with $j^{th}$ ES ($E_j$). In our work, since we consider only the computation demands of the AVs, from now on, we consider ESs and BSs the same and use them interchangeably throughout the text. Our work takes into account the following key considerations:

\begin{itemize}
  \item Tasks generated by dynamic AVs generally have short execution times, ranging from seconds to a few minutes at most \cite{Liu21, Ku21}.
  \item These tasks also come with short deadlines \cite{Ali20}.
  \item The ESs located geographically close to the point of task generation execute the tasks \cite{Dai22, tang}. Each task, if served, is handled by only one ES.
\end{itemize}

\subsection{Task Model}
Autonomous vehicles utilize sensors, including cameras, LiDAR, and various other sensor technologies, to execute critical computational tasks such as object detection, barricade detection, lane detection, and traffic density prediction \cite{Liu21, Ku21}. The tasks are executed concurrently, with one running on the onboard computational unit of the AV and the other on remote ESs. While local execution yields results with low latency, it can occasionally produce inappropriate outcomes due to the limited computational capacity of the onboard unit. In contrast, tasks performed on remote servers benefit from more excellent computational resources, enhancing the accuracy and reliability of the results \cite{Dai22}. We examine two versions of the problem: (a) the offline case, which represents a simplified scenario where the tasks and their corresponding requirements can be precisely predicted in advance. In this context, the scheduler can plan actions accordingly, serving primarily as a theoretical concept for comparative analysis. (b) The online version presents a more realistic approach, wherein some predictions of incoming task data are available beforehand. In this case, the actual data may vary in real-world applications, prompting the scheduler to develop plans based on these predictions and dynamically manage situations as tasks arrive in real time.


We consider $\boldsymbol{I}$ be the set of all tasks in the VES. We uniquely identify each task as $\tau_i$ where $1 \ \leq i \leq N$. All the $M$ ESs collectively receive $N$ tasks altogether. Let the number of tasks received by $ES_j$ be $N_j$. So, we consider $N_1 + N_2 + \cdots + N_j + \cdots + N_M = N$.

We consider that $T_{total}$ as the maximum number of time slots. Let $N^t_j$ be the total tasks that are incoming to $ES_j$ at timeslot $t$. Therefore, we can express $N_j$ as the sum of tasks of overall time slots, written as $N_j = N^1_j+N^2_j+ \cdots + N^t_j + \cdots + N^{T_{total}}_j$. Our aim is to solve an optimization problem for the VES (refer to \autoref{probst}). Although the solution depends on all the ESs, we consider there is no resource transfer between them. The tasks and the solar energy of each ES are indigenous. We apply our approach to each ES independently to generate the final result and then integrate it to get the final solutions. Therefore, we consider solving the problem for a single ES in the system from this point forward.

 We assume an AV ($v_i$) directly connects to a particular ES. Each task ($\tau_i$) is characterized by its arrival time ($a_i$), task generating VU ($v_i$), execution time ($e_i$), deadline ($d_i$), utilization rate ($u_i$), revenue ($r_i$) and is represented by a tuple $\tau_i (a_i, v_i, e_i, d_i, u_i, r_i)$. The utilization rate \( u_i \) of task \( \tau_i \), where \( 0 < u_i \leq 1 \) represents the fraction of a core's capacity consumed by the task. Multiple tasks can share a single core, provided their combined utilization does not exceed the core's total capacity. However, each task can only be executed on one core.

For simplicity, we consider the deadline ($d_i$) of all the tasks to be $a_i + e_i + k$, where $a_i$ denotes the task arrival time, $e_i$ represents the task execution time, and $k$ is an integer ranging from 0 to 5. The value of $k$ is different for different tasks. We also consider that execution time is equal for all tasks, hence we write $e_i=1$. Upon the task's arrival, the VES retains the flexibility to execute the task immediately, defer its execution to any time slot within the deadline, or discard the task altogether. The AVs pay a tentative monthly or yearly fee to the VES to rent its computational resources. The revenue per task ($r_i$) is calculated by dividing the tentative fee by the average VES resource usage. The actual fee is settled later based on the success or failure of the renting process for the tasks over time. So, we approximate the value of $r_i$, which the AV pays the VES upon the successful completion of the task within the deadline. The revenue of task $r_i$ is dependent on $u_i$ and $d_i$ as stated in \autoref{profit formula}. The revenue model is taken from \cite{Mei19, Swain20}.
\begin{equation}
\label{profit formula}
r_i = \frac{{u_i^2}}{{1 + (d_i-a_i)^2}}
\end{equation}

\subsection{Edge Server Model}
\label{prev_sec}

Each ES is associated with a fixed amount of computation power utilization ($U^{canbe}_t$) at a particular time slot ($t$). It basically depends on the total electric power available $P^{avail}_t$ at any particular ES at timeslot $t$. $P^{avail}_t$ is the available eclectic power at time $t$, and the ES uses this power to run computation up to $U^{canbe}_t$. All ESs have a maximum amount of computational power utilization capacity ($U^{max}$), and the ES consumes ($P^{max}$) electric power to operate at its maximum potential. The total amount of computational power utilization by all the tasks in $ES$ at a particular instance of time $t$, i.e. $U_t$, should not exceed $U^{canbe}_t$. The ES might use a portion of the electric power available and store a portion in a battery for future use. $P_t$ is instantaneous power consumption at time $t$ and it corresponds to current utilization of $U_t$. Both $U^{canbe}_t$ and $P_t$ should also not exceed $U^{max}$ and $P^{max}$ respectively. These relations are expressed in \autoref{computationalutilization} and \autoref{powerutilization} respectively. We define a scheduling indicator variable $x^i_{t}$. The value of the variable $x^i_{t} = 1$ if the task $\tau_{i}$ is executed at time slot $t$, 0 otherwise. \autoref{eqn:p1x} states that each task is executed in at most one timeslot.

\begin{equation}
\label{computationalutilization}
    U_t = \sum_{i=1}^N x^i_{t} \cdot u_{i} \leq U^{\text{canbe}}_t \leq U^{max}, \quad \forall t
\end{equation}
\begin{equation}
    \label{powerutilization}
    P_t \leq P^{max},     \:\:\:\: \forall t
\end{equation}
    
\begin{equation}
\label{eqn:p1x}
\sum_{t=1}^{T_{total}}  x^i_{t} \leq 1, \:\:\:\: \forall i 
\end{equation}
where $T_{total}$ is the maximum considered time slots. In this problem, we consider $1 \leq t \leq T_{total}$.

The power consumption of a server is the cubic power of computational power utilization, as defined in Dayarathna et al. \cite{Dayarathna16}. It can be written as:

\begin{equation}
    \label{eqn:power}
    P_{t} = [P_{s} + (P^{max} - P_{s}) . (U_{t} /  U^{max})^3] = F_P(U_t)
\end{equation}
where $P_{t}$ is the power consumed at timeslot $t$,  $P^{max}$ is the maximum solar power, $P_s$ is the edge server's static power requirement when turned on, $U^{max}$ is the maximum utilization available at any time slot. The function $F_P$ calculates the power usage for any given computational usage.

To illustrate, suppose the total number of cores of an ES is 64 ($U^{max}$), and at that point, the power requirement ($P^{max}$) is 2000 units (W). If, at a given time slot $t$, the available power is 1500 units (W) ($P^{avail}_t$), then the ES can operate only 56 cores ($U^{canbe}_t$). However, due to less workload the ES only uses 44 cores (so, $U_t=44$), and $P_t = 920$ W. $P_t$ represents the power consumption to use $U_t$ cores. The remaining power of $1500-920 = 580$ units (W) of power is stored in the battery.

\subsection{Energy Source Model}


\begin{figure}[tb!]
    \centering
    \includegraphics[scale=1]{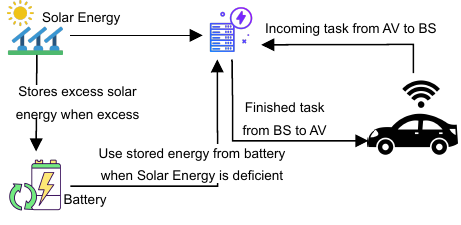}
    \caption{Power model of the considered edge-cloud vehicular computing system}
    \label{powermodelfig}
\end{figure}

 In our case, we considered that solar energy supplies power to the edge-cloud system to make it eco-friendly. \autoref{powermodelfig} shows the energy source model of the proposed system. The BS consists of four units- (a) Edge Server unit- which takes care of the computation demands of the AVs, (b) Solar unit- it collects energy from the sun, (c) Battery- it stores the solar energy and also provides the electric power to provide energy for ESs (d) RSUs- it takes care of the connection between BSs and AVs. This innovative approach promotes sustainability and enhances the system's resilience by providing stored energy for future use.

 In our network, the solar energy ($S_{t}$) contributes to the total electric power ($P_t$) required by the ES. An ES receives $S_1, S_2, \cdots S_t, \cdots , S_{T_{total}}$ amount of solar energy at time instance $T_1, T_2, \cdots , T_t, \cdots , T_{total}$. If at any instance of time instance $t$, the solar energy ($S_{t}$) is surplus to the momentary power consumption requirement $P_t$, the excess power is stored in the battery ($B$). On the other hand, if the power consumption requirement $P_t$ is more than $S_{t}$, the extra power requirement may be provided by the battery $B$ if available. The variable $\beta_{t}$ represents the extra available power stored in $B$ until $t$. The energy model of the proposed system is expressed in \autoref{powermodel_}.

\begin{equation}
    P^{avail}_t = S_t + \beta_t
    \label{powermodel_}
\end{equation}
where $P^{avail}_t$ is total power available at timeslot $t$.

Since the power storage capacity is finite, we assume in \autoref{powermodel_solar}.
\begin{equation}
    \label{powermodel_solar}
    0 \leq S_{t} \leq P^{max}, \:\:\:\:  \forall t
\end{equation}
where $P^{max}$ represents the power required by Edge Servers (ESs) to operate at maximum capacity ($U^{max}$).
 $\beta^{max}$ represents the maximum power storage capacity of $B$. Similarly, the range for battery power is shown in \autoref{powermodel_battery}.

\begin{equation}
    \label{powermodel_battery}
    0 \leq \beta_{t} \leq \beta^{max},\:\:\:\:  \forall t
\end{equation}

\begin{equation}
    U^{canbe}_t = \min(U^{max}, F^{-1}_P(S_t + \beta_t))
\end{equation}
ESs utilization cannot exceed $U^{max}$ and is calculated using \autoref{eqn:power}. $F_P^{-1}$ denotes the reverse function of $F_P$ (\autoref{eqn:power}). It calculates the corresponding computation usage for any given power consumption.

\subsection{Problem Statement}
\label{probst}

The goal of the problem is to maximize the total revenue earned by the VES. In order to do that, we need to maximize the total revenue earned by the VES. Each of the ES receives the solar energy $S_1, S_2, \cdots , S_t, \cdots, S_{T_{total}}$ at time slots $T_1, T_2, \cdots , T_t, \cdots , T_{total}$. Each ES independently uses a portion of this solar power to execute incoming tasks or stores it in a battery for future use, aiming to maximize total revenue. Hence, we formally represent the problem in \autoref{eqn:ps}.
\begin{equation}
    \max \sum_{j=1}^M \left[ R = \sum_{t=1}^{T_{total}} \sum_{i=1}^N  \left( x^i_{t} . r_i \right) \right]
    \label{eqn:ps}
\end{equation}
such that is satisfies \autoref{computationalutilization}, \autoref{powerutilization}, \autoref{eqn:p1x}, \autoref{powermodel_}, \autoref{powermodel_solar}, \autoref{powermodel_battery}.

We consider maximizing the revenue for each ES individually and then integrating them to get maximum revenue for the whole VES.
In our work, we considered various variations of our problem varying the value of $\beta^{max}$, which are listed as follows.
\subsubsection{Task Scheduling on Edge server with Solar Energy and No Battery Backup (TS-ES-SNB)}

It is the simplest version of the problem: solar energy is the only energy source. We consider the value of $\beta^{max}=0$ for this case. 

\subsubsection{Task Scheduling on Edge server with Solar Energy and Infinite Battery Backup (TS-ES-SIB)}
In this version of the problem, the capacity of the battery is infinite, i.e. $\beta^{max} = \infty $. To implement an unlimited capacity battery in a feasible manner, it is assumed that the VES sells any excess solar power back to the electricity distribution company during times of surplus and repurchases the equivalent amount of electric power when needed at every time instance. In this case, we assume that the buying and selling costs are the same, and there is no power loss or power transfer cost.

\subsubsection{Task Scheduling on Edge server with Solar Energy and Finite Battery Backup (TS-ES-SFB)}
Unlike the previous version of the problem, the values of $\beta^{max}$ values are fixed to a finite value.
 \begin{table}[tb!]
 \setlength{\tabcolsep}{0.7mm}
    \centering
    \footnotesize
    \begin{tabular}{|p{2.8cm}|p{5.9cm}|}
      \hline
        \textbf{Notation} & \textbf{Definition}  \\
        \hline
        $\boldsymbol{I}$, $\boldsymbol{E}$, $\boldsymbol{B}$ & Task, edge server, battery set \\
        \hline
         
        $T_{total}$, $\beta^{max}$ & Maximum time, battery storage \\
        \hline
        $U^{canbe}_t$,  & Total utilization available based on available \\        
        $P^{avail}_t$ &  power, available power at time $t$ \\
        \hline
        $P_t$, $U_{t}$ & Power consumption and current utilization at $t$ \\
        \hline

        $U^{canbe_{pred}}_t$, $U^{canbe_{actual}}_t$ & Predicted utilization, actual utilization of task  \\
        \hline

        $S_t$,$\beta_{t}$ & Solar energy available at $t$, battery storage till $t$ \\
        \hline
       
        $S_t^{pred}$, $S_t^{actual}$ & Predicted solar power, actual solar power \\
        \hline
        $P_s$, $U^{max}$, $P^{max}$ & Static power, max utilization, max power\\
        \hline

        $P_{save}$, $P_{move}$ & Power saved by dropping tasks, power shifted \\
        \hline
        $TQ$, $R$ & Task Queue, Total Revenue \\
        \hline
        $\tau_{drop}$, $\tau_{exe}$ &  Task which is to be dropped, executed\\
        \hline
        $t_{max}$, $t_{min}$ & Timeslot with max and min power consumption\\
        \hline
        $P_{save}$, $P_{move}$, $P_{avg}$ & Amount of power saved by task drop from $t_{max}$, amount of power moved from $t_{max}$ to $t_{min}$, average power consumption across all timeslots\\
        \hline
        $P_d$, $T_d$ & Deviation of solar profile and task profile from predicted data\\
        \hline
        $I^{pred}_t$, $I^{actual}_t$ & Predicted task set, actual task set at time $t$\\
        \hline
        $I^{pred_{scheduled}}_t$, $I^{actual_{offlinescheduled}}_t$ & 
        Tasks from $I^{pred}_t$ that is scheduled, tasks from $I^{pred_{scheduled}}_t$ that arrive actually in ES\\
        \hline
        \(N_t^{TQ}\), \(N_t^{exe}\) & The number of tasks in the $TQ$ at time \(t\), the number of tasks from the predictions provided currently executing in the ES at time $t$.\\
        \hline

    \end{tabular}
    \caption{Notations used}
    \label{Table_1}
\end{table}

Our problem is particularly challenging because the power usage increases cubically with computational usage. This requires careful decision-making about which tasks to execute and at what timeslots within their deadlines they should execute. Additionally, we must determine when and how much solar power to store for future use, especially given the limited battery storage capacity. The complexity increases further when the predicted values for task demands and solar power differ from the actual values. These factors collectively present significant challenges for our problem.

\section{Solutions Approach}
In our work, we address three primary problems, offering solutions for each. For the first problem, TS-ES-SNB, we present an optimal solution for a specific case of the problem and a heuristic solution for the general case. The heuristic solution is detailed in \autoref{alg:S1}. It is solved assuming an accurately predicted task arrival pattern and solar profile for each time slot. It is an offline approach because it executes on data known initially.

For the second problem, TS-ES-SIB, we propose heuristics for both offline and online scenarios. This solution for offline scenarios is generated in advance using predicted data. The solution for the offline scenario leverages predicted vehicular task data and predicted solar profiles, combined with the general solution of TS-ES-SNB, to formulate the solution. The corresponding pseudocode is shown in \autoref{alg:unlimited_battery}. However, in real-world scenarios, the incoming task data and solar profiles may vary to some extent from predicted values. To address this, we propose an online scheduling approach, detailed shown in \autoref{alg:online_scheduling}. It integrates the offline solution with both predicted and real-time data to deliver a robust solution.

\begin{figure}[tb!]
    \centering
    \includegraphics[scale = .61]{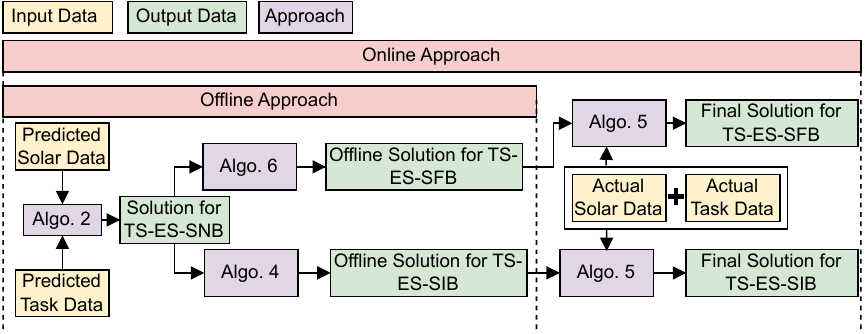}
    \caption{Flowchart of the Proposed Approaches}
    \label{fig:fc}
\end{figure}

Similarly, for the third problem, TS-ES-SFB, we proposed a heuristic approach for offline scenarios, and it is illustrated in \autoref{alg:limited_battery}. This approach utilizes the TS-ES-SNB solution and predicted data to develop an initial solution. For the online scenario of TS-ES-SFB, we again use \autoref{alg:online_scheduling}, which combines the offline solution with real-time data, ensuring adaptability and accuracy in dynamic conditions. \autoref{fig:fc} shows the detailed flowchart of all our approaches.

Autonomous vehicles (AVs) offload the task $\tau_i$ to nearby BSs equipped with edge server $E_j$. Tasks arrive in batches, and each task must be executed before the deadline or dropped. The edge servers decide upon which task to execute and when based on available solar power and with the objective of revenue maximization. In the offline scenario, the task profile across all the timeslots and energy profile across all the timeslots $S_{t}$ are known beforehand. It is beneficial to distribute power consumption across all the timeslots to ensure better utilization of available resources since the power consumption increases cubically with respect to computational resource usage. However, this is challenging due to the fluctuating task demands across different timeslots. Power can only be shifted from earlier to later timeslots because stored energy can be used only if stored.

\subsection{Solution Intuition}

We leverage a distinctive characteristic of the cubic power model to optimise revenue. According to the power model in \autoref{eqn:power}, the model takes utilization as input and outputs the corresponding power consumed. Let \(A\) and \(B\) represent two utilization levels for a specific ES. The power consumed for these cases is \(F_P(A)\) and \(F_P(B)\), respectively. For two integers \(A\) and \(B\) where \(B > A\), it follows that:
$
(B+1)^3 - B^3 > (A+1)^3 - A^3.
$

This indicates that an increase in utilization demand by 1 unit results in a greater increase in power consumption when the initial utilization usage is higher. Consequently, when two new tasks with utilization demand \(C\) arise, and we have limited power in the battery, it is beneficial to schedule the task that comes in the timeslot where the initial utilization demand is less. This is because the increase in power consumption is minimized if it is scheduled during a lower initial power consumption period. Specifically, it holds that:
$
F_P(B+C) - F_P(B) > F_P(A+C) - F_P(A), \: \forall\ A, B, C > 0 \: \text{and} \: B>C 
$.
This condition holds since $A$, $B$, and $C$ are always positive in practical scenarios.

\begin{figure}
    \centering
    \includegraphics[scale =0.8]{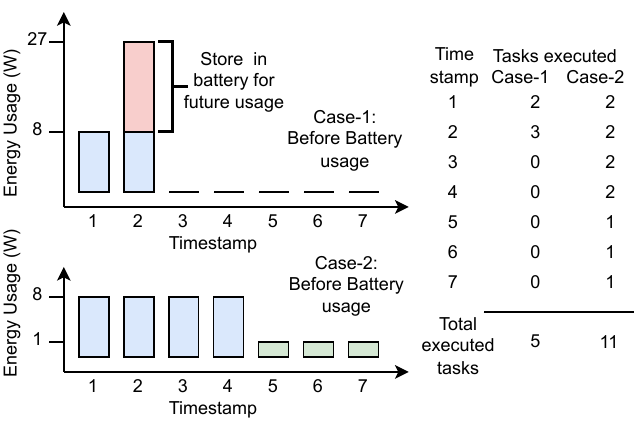}
    \caption{A simple example to demonstrate why redistribution of power is beneficial. The power model is used in \autoref{eqn:power}. For ease of calculation, we considered that the values in$P_s=0$, $U^{max}=1$, $P^{max}=1$ and all tasks with utilization of 1.}
    \label{fig:example}
\end{figure}

With this property in mind, we aim to shift power consumption from periods of higher consumption to those of lower consumption. Solar power is abundant during the day and absent at night, leading to a gradual increase in energy availability throughout the day and peaking at zero at night. This solar energy can be stored in batteries for nighttime use. Thus, redistributing power from high-consumption time slots to low-consumption ones is advantageous. As illustrated in \autoref{fig:example}, redistributing power can enhance the total number of tasks completed while total power consumption is constant. For simplicity in this example, we set parameters \(P_s=0\), \(U^{max}=1\), \(P^{max}=1\), and assumed all tasks have a utilization demand of \(1\) (i.e., \(u_i=1\)). In this scenario, the number of tasks completed by the ES is increased from 5 to 11.

However, this approach poses two challenges: 
(a) Power can only be shifted from past timeslots with higher power consumption to future timeslots with higher power consumption since energy must be stored before use. For simplicity, we assume no power loss occurs during the charging and discharging of the battery. The complexity escalates when battery capacity is limited, restricting the amount of energy that can be stored. 
(b) Task demand fluctuates throughout the day, making it inefficient to transfer power to periods of no demand.

\begin{figure}
    \centering
    \includegraphics[scale =0.5]{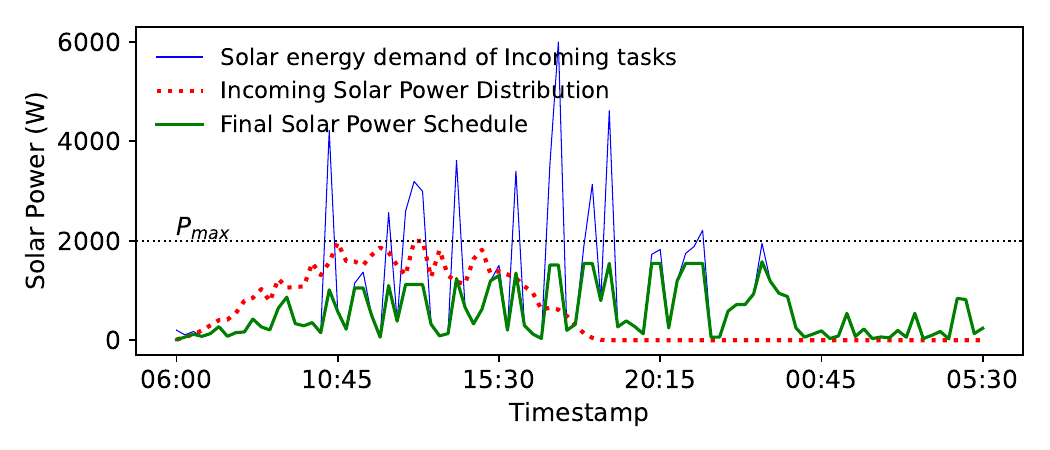}
    \caption{A sample solution using the proposed approach, which increased the number of completed tasks by 44\% with the amount of incoming solar power being constant.}
    \label{fig:sample_soln}
\end{figure}

Our proposed solution addresses both challenges. A sample solution is shown in \autoref{fig:sample_soln}, illustrating how solar power generated during the day is stored in batteries for nighttime use. While the total solar power utilized remains unchanged in both scenarios, the overall revenue increases by 44\%.

\subsection{Solution to Task Scheduling on ES with Solar Energy with no Battery (TS-ES-SNB)}

\subsubsection{Special case :} 
\begin{figure} [tb!]
    \centering
    \includegraphics{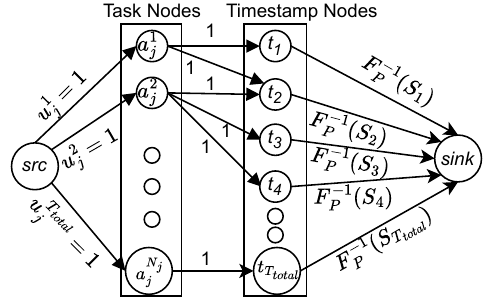}
    \caption{The special case of TS-ES-SNB problem is mapped to Maximum Flow problem \cite{ford_fulkerson_1956} with the nodes and maximum flow capacities mentioned for a particular ES ($ES_j$).}
    \label{fig:specialcase}
\end{figure}

When the computation demand by tasks and revenue earned by VES on successful completion of task within the deadline is 1 for all tasks ( $u_i=1$, $r_i=1$, $\forall i$).
The problem can be mapped to the Maximal Flow through a Network \cite{ford_fulkerson_1956}. Let a directed graph $G = (V, E)$ with a source node ($src$) and sink node ($sink$), where each edge $(x,y)$ has a capacity $c(x,y)$. \autoref{fig:specialcase} shows the special case of the TS-ES-SNB problem mapped to the Maximum Flow problem for a particular ES ($ES_j$).
In our proposed scenario, we establish a mapping to model the system as a flow network. The mapping is as follows:

\begin{enumerate}
    \item {Node Creation:}
    We create nodes to represent each task and timestamp, along with dedicated source ($src$) and sink ($sink$) nodes.

    \item {Source to Task Edges:}
    An edge is added from the source node to each task node, designated as ($src$, $a_j^i$), where $a_j^i$ presents the arrival time for $\tau_i$ at $ES_j$.The edge has a unit capacity because $u^i_j=1$. For example, the task $\tau_1$ arrives at time $a_1$, so there is a edge connecting ($src$, $a_1$).

    \item {Task to Timestamp Edges:}
    For every task, node connections are established to all the timestamp nodes falling within the arrival and deadline interval of task $\tau_i$.
        If a timestamp $t$ falls within this interval, edge $(a_j^i, t)$ with a unit capacity is introduced for all $a_i \leq t < d_i$ for task $\tau_i$. For example, the task $\tau_1$ arrives at time $a_1$ and the deadline $d_1$. So there are edges connecting ($a_1$, $t_1$) and ($a_1$, $t_2$), since we consider $a_1 \leq t_1, t_2 < d_1$. Similarly the task $\tau_2$ arrives at $a_2$ and there is edges connecting ($a_2$, $t_2$), ($a_2$, $t_3$), ($a_2$, $t_4$) since we consider $a_2 \leq t_2, t_3, t_4 < d_2$.

    \item {Timestamp to Sink Edges:}
    Connections are established from each timestamp node $t$ to the sink node, denoted as ($t, sink$).
        The capacity of these edges corresponds to the utilization that can be fueled by solar power at timestamp tt. For example, at time $t_1$, $S_1$ amount of solar power is available, which can support $F_p^{-1}(S_1)$ amount of computation.
    
\end{enumerate}

The objective is to maximize the total flow from $src$ to $sink$, subject to the following constraints:

\begin{enumerate}
    \item {Capacity Constraints:}
    \begin{equation}
    \label{cap}
    0 \leq f(x,y) \leq c(x,y) \quad \forall (x,y) \in E
    \end{equation}
    This constraint ensures that the flow ($|f|$) along each edge does not exceed its capacity.
    
    \item {Flow Conservation Constraints:}
    For each node $y$ other than $src$ and $sink$, the total incoming flow equals the total outgoing flow:
    \begin{equation}
    \label{cons}
    \sum_{x \in V} f(x,y) = \sum_{w \in V} f(y,w) \quad \forall y \in V \setminus \{src,sink\}
    \end{equation}
\end{enumerate}

The goal is to maximize $|f|$, the total flow from $src$ to $sink$ while satisfying the capacity and flow conservation constraints. In our case, max flow $|f|$ is mapped to maximum revenue. This is the same as solving the mentioned special case of the problem TS-ES-SNB. Since the optimal solution for this problem exists (Ford et al. \cite{ford_fulkerson_1956}), so we can solve our problem optimally.

\subsubsection{General case}
For the general case, the value of $u_i$ and $r_i$ may not be 1, the values of $u_i$ and $d_i$ may vary and the value of $d_i$ is calculated as per \autoref{profit formula}. The pseudocode for the heuristic solution of problem TS-ES-SNB is shown in \autoref{alg:S1}. The value of maximum utilization at time $t$ ($U^{canbe}_t$) is calculated from available solar power $S_t$ using \autoref{eqn:power}. In this case $\beta^{max} = 0$ hence $P^{avail}_t = S_t$. 

This approach is for ES with no-battery backup, which means at any timeslot $t$ whatever the available solar power is, we have to use it in that timeslot only, and it cannot be stored in the battery for later use. Hence, the approach schedules tasks by starting from the last timeslot and moving towards the first. At each timeslot, first of all, the tasks $\tau_{i}$, which are available, are pushed into the task queue ($TQ$), and tasks are selected based on the highest $r_{i}$. The selected tasks execute as described in pseudocode of \autoref{alg:te}.
The cumulative utilization of all selected tasks should not exceed the utilization available at time $t$, i.e $U^{canbe}_t$. When the task is executed, the $r_i$ is added to the total R, and the task is removed from task queue TQ.
 This step aims to allocate tasks while efficiently utilizing solar energy.

\begin{algorithm}[tb!]
\footnotesize
    \caption{Procedure Task Execution}
    \label{alg:te}
\KwIn{$\tau_i$, $t$, $U^{canbe}_t$, $U_{t}$, $R$, $TQ$} \KwOut{Updated $R$}

\If{$U_{t} + u_i \leq $$U^{canbe}_t$}
{

Execute($\tau_i$,$t$) ; $U_{t} \gets U_{t} + u_i$; 

Update $R \gets R$ + $r_i$ ; Remove $\tau_i$ from $TQ$
}

Return $R$

\end{algorithm}

\setcounter{AlgoLine}{0}
\begin{algorithm}[tb!]
\footnotesize
    \caption{Schedule Tasks on ES without battery: TS-ES-SNB}
    \label{alg:S1}
\KwIn{$I^{pred}_t$, $T_{total}$, $U^{canbe}_t$, $U_{t}$=0, $R=0$, $TQ$}
\KwOut{Initial Schedule, updated R}

\For{$t$ in range($T_{total}$,1) }{

  \For{all $\tau_i$ with $a_i \leq t $ and $d_i \geq t+e_i $ }{
  Choose $\tau_i$ with maximum ${r_i}$

    R$\gets$Task Execution($\tau_{exe}$,$t$,$U^{canbe}_t$, $U_{t}$,$R$,$TQ$) using \autoref{alg:te}

    }
 
}

\end{algorithm}

\subsection{Solution to Task Scheduling on ES with Solar Energy and Infinite Battery (TS-ES-SIB)}

In this version of the problem, we assume that solar power is equipped with battery storage with infinite capacity ($\beta^{max}=\infty$). This scenario is just an assumed scenario for theoretical analysis and comparison purposes. This gives a better insight into the problem for the actual scenario case, which we use in a later stage.

\setcounter{AlgoLine}{0}

\begin{algorithm}[tb!]
\footnotesize    
    \caption{Procedure Task Dropping}
    \label{alg:td}
\KwIn{$\tau_{drop}$, $t$, $U^{canbe}_t$, $U_{t}$, R, TQ}
\KwOut{Updated $R$, $P_{save}$}

$P_{save} \gets F_P(U_{t}) - F_P(U_{t} - u_{drop})$

Drop($\tau_{drop}$,t); Add $\tau_{drop}$ to TQ

$U_{t} \gets U_{t} - u_{drop}$; Update R $\gets$ R - $r_{drop}$

Return R, $P_{save}$

\end{algorithm}

\setcounter{AlgoLine}{0}

\subsubsection{Offline Scenario}

\begin{algorithm}[tb!]

\footnotesize
\caption{Offline Schedule for Infinite Battery: TS-ES-SIB}
\label{alg:unlimited_battery}

 \KwIn{$I^{pred}_t$, $T_{total}$, $U^{canbe}_t$, $U_{t}$, $R$, $TQ$}
 \KwOut{Final offline schedule, updated R}

Create initial schedule using No Battery (\autoref{alg:S1} ($I^{pred}_t$, $T_{total}$, $U^{canbe}_t$, $U_{t}$=0, $R=0$, $TQ$))
 
\Repeat{($R$ is increasing)}{
    $t_{max} \gets$ timestamp $t$ where $U_{t}$ is maximum

    $\tau_{drop} \gets \tau_i$ scheduled at $t_{max}$ with minimum ${r_i}$

    R,$P_{save}\gets$Task Drop($\tau_{drop}$,$t_{max}$,$U^{canbe}_{{t_{max}}}$,$U_{{t_{max}}}$,R,TQ) using \autoref{alg:td} 

    \While{$P_{save} > 0$}{

    \For{$ t_{max}< t \leq T_{total}$}{
    $t_{min} \gets$ timestamp $t$ where $U_{t}$ is minimum }

    $\tau_{exe}\gets$ Choose $\tau_i$ with maximum ${r_i}$ at time $t_{min}$

    R$\gets$Task Execution($\tau_{exe}$,$t_{min}$,$U^{canbe}_{{t_{min}}}$, $U_{{t_{min}}}$,$R$,$TQ$) using \autoref{alg:te}

    \If{$\tau_{exe}$ is scheduled}{

    $P_{save} \gets P_{save} - (F_P(U_{{t_{min}}}+u_{exe}) - F_P(U_{{t_{min}}}))$
   
    }
  
    }
  
    }

\end{algorithm}

We focus on balancing the load in terms of utilization across all the timeslots to optimize the schedule further. This approach uses the initial solution results generated by TS-ES-SNB (\autoref{alg:S1}). The heuristic solution's pseudocode for the TS-ES-SIB offline problem is shown in \autoref{alg:unlimited_battery}. The approach iteratively identifies the most and least consumed timeslots in terms of utilization, i.e., current utilization $U_{t}$.

Here we identify two time slots $t_{max}$ and $t_{min}$. The power used in $t_{max}$ is the highest in the entire schedule, and power used at time $t_{min}$ is the lowest in the entire schedule, and $t_{min}>t_{max}$ on timeline scale. At the maximum power consumed timeslot $t_{max}$, a task with minimum revenue $r_i$ is dropped (pseudocode as given in \autoref{alg:td}) and when the task is dropped at timeslot $t_{max}$, then  $P_{save}$ is the power saved at that timeslot $t_{max}$, which is to be stored in the battery. The dropped task is reconsidered for execution in other time slots between its arrival and deadline if it is found to be more profitable than other tasks. Consequently, the dropped task is re-entered into the task queue ($TQ$), and the total revenue $R$ is updated accordingly. With the excess power $P_{save}$ at $t_{min}$ now, we can execute more tasks at $t_{min}$. At timeslot $t_{min}$ from all unallocated tasks which are available at timeslot $t_{min}$, the task $\tau_{exe}$ with maximum $r_i$ is selected and task executes (as described in \autoref{alg:te}). If there exists a task that can be scheduled satisfying utilization constraint given by \autoref{computationalutilization}, then $P_{save}$ gets updated as 

$P_{save} \gets P_{save} - (F_P(U_{{t_{min}}}+u_{exe}) - F_P(U_{{t_{min}}}))$

where $F_P(U_{{t_{min}}})$ is consumed power at timeslot $t_{min}$ before executing the task $\tau_{exe}$, $F_P(U_{{t_{min}}}+u_{exe})$ is consumed power at $t_{min}$ after executing $\tau_{exe}$ their difference is subtracted from $P_{save}$.

Tasks are selected and executed iteratively till $P_{save}$ is positive. The approach first searches for $t_{max}$ and transfers power to a bunch of $t_{min}$. This is because the power from a timeslot can be shifted to multiple timeslots. It again identifies another $t_{max}$ and continues the same process until the value of total revenue earned ($R$) increases.

\subsubsection{Online Secnario}

\setcounter{AlgoLine}{0}

\begin{algorithm} [tb!]

\footnotesize

\caption{Online Scheduling Approach}
\label{alg:online_scheduling} 

\KwIn{$I^{pred}_t$, $I^{actual}_t$, $T_{total}$, $U^{canbe}_t$, $U_{t}$, $R$, $TQ$}
\KwOut{Final online schedule, Updated R}

Generate offline scheduling using \autoref{alg:unlimited_battery} or \autoref{alg:limited_battery} ($I^{pred}_t$, $T_{total}$, $U^{canbe}_t$, $U_{t}$, $R$, $TQ$) depending on battery capacity $\beta^{max}$

\For{$t$ in range (1, $T_{total}$)}
{$U_t$ from offline predicted data is used as input for online.

\eIf{$F_P(I^{actual_{offlinescheduled}}_t) > S^{actual}_t$}{
Drop tasks which scheduled at $t$ with minimum $r_i$ and add to TQ till $F_P(I^{actual_{offlinescheduled}}_t)$ is equal to $ S^{actual}_t$\\

}
{
Schedule tasks from TQ at $t$ with maximum $r_i$ till $F_P(I^{actual_{offlinescheduled}}_t)$ is equal to $ S^{actual}_t$\\

}
Update $R$
}
\end{algorithm}

In this scenario, the exact task profile and solar energy profile throughout timeslot $T_{total}$ is unknown. $I^{actual}_t$ represents the actual task profile at timeslot $t$ and  $I^{pred}_t$ represents the predicted task profile at timeslot $t$ respectively. However, the batch of tasks coming at different timeslots and the nature of these tasks can be predicted with some error. Similarly, the solar energy availability at different timeslots can be approximated based on historical data with some errors. 

However, in a real-life scenario, the number and nature of incoming tasks may not precisely match the predictions, and solar power may deviate from the predicted values. In our study, we considered variations of 5\% and 10\% between the predicted and actual data for solar power and incoming task sets. We use offline schedule derived using \autoref{alg:unlimited_battery} as an input to solution approach to the online scenario.

$S_t^{pred}$ represents the predicted solar profile at time $t$ and $U^{canbe_{pred}}_t$ represents the computation usage at time $t$ from scheduled solar profile respectively. $S_t^{actual}$ represents the actual solar profile at time $t$ and $U^{canbe_{actual}}_t$ represents the actual computation usage at time $t$ respectively.  

The approach for offline scenario uses $I^{pred}_t$ for incoming task data; a subset of it ($I^{pred_{scheduled}}_t$) is scheduled. The rest remain unscheduled. It might happen that some tasks predicted from $I^{pred_{scheduled}}_t$ do not arrive at the ES, so our approach for the online scenario chooses the most profitable tasks for unscheduled tasks or new tasks and maximizes its revenue. We define the scheduled tasks and arrive at ES as $I^{actual_{offlinescheduled}}_t$.

There can be two cases between $I^{actual_{offlinescheduled}}_t$ and $S_t^{actual}$.
\begin{itemize}
    \item {Case 1:} $F_P(I^{actual_{offlinescheduled}}_t) > S^{actual}_t$: If the actual solar power is lesser than the required solar power ($F_P(I^{actual_{offlinescheduled}}_t)$). In that case, we need to drop some tasks from $F_P(I^{actual_{offlinescheduled}}_t)$. The tasks that are dropped are with minimum $r_i$, and it continues till $F_P(I^{actual_{offlinescheduled}}_t)$ becomes equal to $ S^{actual}_t$.
    
    \item {Case 2:} $F_P(I^{pred_{actual_{offlinescheduled}}}_t) \leq S^{actual}_t$: If the actual solar power is more than the required solar power, then we add tasks for execution on ES from $TQ$ which are available at that moment. We select tasks from TQ with maximum $r_i$. The task execution continues till $F_P(I^{actual_{offlinescheduled}}_t)$ becomes equal to $ S^{actual}_t$.

\end{itemize}

\subsection{Solution to Task Scheduling on ES with Solar Energy and Finite Battery (TS-ES-SFB)}\label{sub_section_sfb}

\subsubsection{Offline Secnario} 

In the finite battery case, the capacity of the battery is limited and is given by  $\beta^{max}$. Unlike the previous problem TS-ES-SIB, suppose we aim to store power in the battery to shift from the $t_1$ to the $t_4$ time slot; we cannot use that storage to shift power from the $t_2$ to the $t_3$ time slot, where $t_1 < t_2 < t_3 < t_4$. So, we need a different approach to solve this problem.

The input to this approach is the output of the TS-ES-SNB problem (of \autoref{alg:S1}). The pseudocode of this proposed approach is shown in \autoref{alg:limited_battery}, where we strive to balance power consumption across each timeslot. Given the finite battery capacity, we focus on short-term storage to facilitate greater power transfer. The aim is to maintain power consumption in each timeslot close to the average consumption level, denoted as $P_{avg}$.

The approach goes through an iterative process to identify two timeslots: firstly, the timeslot $t_{max}$ where the power consumption $F_P(U_{{t_{max}}})$ is maximum, and secondly, a subsequent timeslot $t_{min}$ which is the immediate next timeslot (on timeline scale) to $t_{max}$ where the power consumption $F_P(U_{{t_{min}}})$ is below the average consumption $P_{avg}$. For example, timeslot $t_{max+5}$ and timeslot $t_{max+10}$ have power consumption less than $P_{avg}$. Additionally all time slots in range ($t_{max+1}$ to $t_{max+4}$ and $t_{max+6}$ to $t_{max+10}$) have power consumption more than $P_{avg}$. So, $t_{max+5}$ is selected as $t_{min}$ to transfer power as it is nearer to $t_{max}$.

The battery stores and transfers power from $t_{max}$ to $t_{min}$, creating small transfer zones in the time axis that facilitate more efficient power transfers since the battery is engaged for shorter durations. The amount of power $P_{move}$ that can be transferred (from one time slot in past to future time slot) depends on a minimum of three factors: the available battery capacity at time $t$ ($\beta^{max}-\beta_t$), the difference between the power consumed at $t_{max}$ i.e. $F_P(U_{{t_{max}}})$ and $P_{avg}$, and the difference between $P_{avg}$ and the power consumed at $t_{min}$ i.e $F_P(U_{{t_{min}}})$.

The transferred power $P_{move}$ is stored in the battery, and tasks are dropped based on minimum $r_i$ at timeslot $t_{max}$. At $t_{min}$, from a pool of unscheduled tasks in the task queue, the algorithm selects the task available at $t_{min}$ based on the maximum $r_i$. This dropping process and scheduling continue iteratively until the saved power after the task drops $P_{save}$ equals $P_{move}$. 

\setcounter{AlgoLine}{0}
\begin{algorithm}[tb!]
\footnotesize
 \caption{Offline Schedule for Fixed Battery : TS-ES-SFB}

 \label{alg:limited_battery}
 \KwIn{$I^{pred}_t$, $T_{total}$, $U^{canbe}_t$, $U_{t}$, $R$, $TQ$}
 \KwOut{Final offline schedule, updated $R$}

 Create initial schedule using No Battery (\autoref{alg:S1}($I^{pred}_t$, $T_{total}$, $U^{canbe}_t$, $U_{t}$=0, $R=0$, $TQ$)) 
 
 Average power consumption ($P_{avg}$)$\gets 
  \frac{\sum_{t=1}^{T_{total}} F_P(U_{t})}{T_{total}}$

 \Repeat{($R$ is increasing)}{
 $t_{max} \gets$ timestamp $t$ where $F_P(U_{t})$  is maximum

 \For{t in range($t_{max}+1$,$T_{total}$)}{
\If{$F_P(U_{t}) < P_{avg}$}{
$t_{min} \gets t$; break
}
 
 }

 $P_{move} \gets min(\beta^{max}- \beta_t ,F_P(U_{{t_{max}}})-P_{avg},P_{avg}-F_P(U_{{t_{min}}}))$

\While{$P_{move} > 0$}{

$\tau_{drop} \gets \tau_i$ scheduled at $t_{max}$ with minimum ${r_i}$
    
    R,$P_{save}\gets$Task Drop($\tau_{drop}$,$t_{max}$,$U^{canbe}_{{t_{max}}}$,$U_{{t_{max}}}$,R,TQ) using \autoref{alg:td}

    \While{$P_{save} > 0$}{

    $\tau_{exe}\gets$ Choose $\tau_i$ with maximum ${r_i}$ at time $t_{min}$

    R$\gets$Task Execution($\tau_{exe}$,$t_{min}$,$U^{canbe}_{{t_{min}}}$, $U_{{t_{min}}}$,R,TQ) using \autoref{alg:te}

    \If{$\tau_{exe}$ is scheduled}{

    $P_{save}\gets P_{save} - F_P(U_{{t_{min}}}+u_{exe}) - F_P(U_{{t_{min}}})$

    }

    }

    $P_{move} \gets P_{move} - P_{save}$
}
 
 }

\end{algorithm}

\subsubsection{Online Secnario} 
In this scenario, also as described in the online approach for unlimited battery case (TS-ES-SIB), the exact task profile
$I$ and solar energy profile throughout timeslots is not known. The steps are the same as described in the online approach unlimited battery case except for the power shifting phase. Here, we have finite battery capacity; hence, we use the approach shown in \autoref{alg:limited_battery} (for TS-ES-SFB) for load balancing. The online part is solved similarly to the approach given in \autoref{alg:online_scheduling}, described earlier.

\subsection{Scheduler Overhead}

\begin{figure}
    \centering
    \includegraphics[scale =0.8]{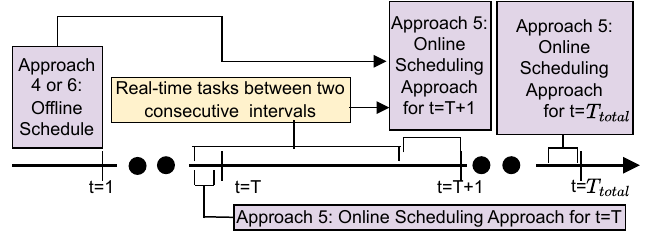}
    \caption{Timeline for approaches activation}
    \label{fig:algotimeline}
\end{figure}

In \autoref{fig:algotimeline}, we observe that the offline scheduling solution is determined prior to the commencement of real-time operations, utilizing predicted data. We use the approach in \autoref{alg:unlimited_battery} to address scenarios with an infinite battery supply and the approach in \autoref{alg:limited_battery} for those with a finite battery. During real-time execution, we integrate the offline solution with real-time power availability and incoming task data to construct the schedule for the online case using the approach given in \autoref{alg:online_scheduling}. In this online case, we consider tasks and power data from the previous batch's onset to the current batch's start. There is a delay in generating the real-time schedule, and task execution commence at specific points, as indicated in \autoref{fig:algotimeline} marked with $t$.

The time complexity of the online approach can be calculated as:

\[
O\left((N_t^{TQ} + N_t^{exe}) \left( \log(N_t^{TQ}) + \log(N_t^{exe}) \right)\right)
\], 
at time \(t\), where \(N_t^{TQ}\) represents the number of tasks in the task queue at time \(t\), and \(N_t^{exe}\) denotes the number of tasks from the predictions provided currently executing in the ES at time $t$. The expected value of the number of tasks in the queue is calculated as:
$E(N_t^{TQ}) = \frac{\sum (N \cdot E(k))}{T_{total}}
$ and the expected value of the number of tasks executing is calculated as $E(N_t^{exe}) \:=\:
\frac{\sum_{t=1}^{T_{total}} x_t^i}{T_{total}}
$.

If, at a given time \(t\), the power required by tasks scheduled through the offline approach exceeds the actual available power, i.e., 
$F_P(I^{actual_{offlinescheduled}}_t) > S^{actual}_t$, 
the time complexity of the online approach was reduced to:

\[
O\left(N_t^{exe} \left( \log(N_t^{TQ}) + \log(N_t^{exe}) \right)\right)
\]

, as we must drop the task with the minimum \(r_i\). A min-heap is utilized to manage the executing tasks from the predictions, while a max-heap is maintained for the real-time tasks that arrive in the queue. Extracting the task with the minimum \(r_i\) from the executing predicted tasks at time \(t\) requires \(O(\log(N_t^{exe}))\), while extracting the task with the maximum \(r_i\) from the task queue takes \(O(\log(N_t^{TQ}))\), followed by an addition to the executing tasks heap that also takes \(O(\log(N_t^{exe}))\). Given that this operation can occur up to \(N_t^{exe}\) times for each time \(t\), the time complexity for this scenario is:

\[
O\left(N_t^{exe} \left(\log(N_t^{TQ}) + 2\log(N_t^{exe})\right)\right) = 
\]
\[
O\left(N_t^{exe} \left(\log(N_t^{TQ}) + \log(N_t^{exe})\right)\right).
\]

In the other case, the power required by tasks scheduled through the offline approach does not exceed the actual available power, i.e., 
$F_P(I^{actual_{offlinescheduled}}_t) \leq S^{actual}_t$,  the time complexity is given by:

\[
O\left(N_t^{TQ} \left(\log(N_t^{TQ}) + \log(N_t^{exe})\right)\right).
\]

Here, extracting the task with the maximum \(r_i\) from the task queue takes \(O(\log(N_t^{TQ}))\), and adding it to the heap of executing predicted tasks takes \(O(\log(N_t^{exe}))\). This operation can occur a maximum of \(N_t^{TQ}\) times for each time \(t\), leading to the complexity.

\[
O\left(N_t^{TQ} \left(\log(N_t^{TQ}) + \log(N_t^{exe})\right)\right).
\]

Thus, the overall time complexity of the online scheduling approach is:

\[
O\left(N_t^{exe} \left(\log(N_t^{TQ}) + \log(N_t^{exe})\right)\right)\]
\[
 + O\left(N_t^{TQ} \left(\log(N_t^{TQ}) + \log(N_t^{exe})\right)\right) =\]
\[
O\left((N_t^{TQ} + N_t^{exe}) \left(\log(N_t^{TQ}) + \log(N_t^{exe})\right)\right).
\]

\section{Experiments and Results}

We developed a homegrown simulator using C++ and Python to simulate VES with solar energy and battery storage. We compared the results of our approach with various state-of-the-art (SOTA) approaches in real-life datasets.

\subsection{Real-life Dataset}

We used the approach given in Tang et al. \cite{tang} to get the traffic flows during different times of the day. As computational requirements and deadlines are not available in the dataset, and there is no method to generate the computation requirements of the AVs and the deadline of the computation requirements from any real-life datasets, we considered the computation requirements using Gaussian distributions. We use the term $\rho$ as the ratio of average $u_i$ and average $(d_i - a_i)$ of all the tasks. We chose five values of $\rho$ and experimented on these datasets. We considered a task trace of 100 K spread over 24 hours for each of the values of $\rho$ and ran experiments on the dataset. As the value of $\rho$ increases, the difficulty level to schedule the task also increases as the utilization requirements of the task increase and the deadline decreases. The revenue associated with the task with higher $\rho$ is higher compared to lower values of $\rho$.

We used the dataset of Annal et al. \cite{solar_power_generation_data} to export the solar energy profile. We collected solar data from the dataset for each minute and used it to run our experiments. We normalized the power data within a scale between 0 and 2000 W to ease our calculation.

It is challenging to predict task profiles in VES accurately. This is because of the changing traffic dynamic due to factors like traffic load, weather, driver efficiency etc. Tang et al. \cite{tang} states that the variation of residency time of an AV within the range of an ES can vary as much as 20 minutes. We used the FUSD approach as mentioned in Xiao et al. \cite{onlineprocedure} to get the real-time solar profile data from predicted data. We define the term $P_d$ where $P_d$ indicates $S_t^{actual}$ deviates from $S_t^{pred}$ by $d$\%. So, $S_t^{actual}$ can be any values between $\frac{100-d}{100} * S_t^{pred}$ to $\frac{100+d}{100} * S_t^{pred}$. We used the $S_t^{actual}$ as the base and used the FUSD approach to get $S_t^{pred}$. $S_t^{actual}$ increases quickly compared to $S_t^{pred}$ when the amount of solar energy increases and decreases slowly when the amount of solar energy decreases. The term $T_d$ indicates that $100-d$\% of task from predicted task data arrive at the VES, while the rest of the tasks ($d\%$) either do not arrive or arrive with different values of $u_i$, $d_i$ and $r_i$ arrive instead of those tasks.

We conducted different solution approaches to the proposed problem by fixing the following parameter values: $P_{max} = 2000$, $U_{max} = 100$, $P_s = 20$ which corresponds to real-life values as claimed in \cite{Dayarathna16}. 

\subsection{State-of-the-art methods}

We implemented the state-of-the-art YARN scheduler \cite{yarn}, which is a global scheduling framework. However, it proved unsuitable for our scenario, as task migration is not a factor. Our primary focus is on efficiently storing and utilizing energy for future use. We adapted the YARN scheduler to fit our needs; it performed sub-optimally. YARN utilizes fairness and capacity scheduling mechanisms, but fairness does not apply to our work because energy availability varies across different locations and times. Our objective is to minimize energy wastage.

In terms of capacity, YARN's capacity scheduler aims to reduce resource usage while preventing hotspots, focusing mainly on minimizing static energy consumption. In contrast, our approach prioritizes minimizing dynamic energy consumption, as energy usage scales cubically with computational load. Furthermore, we aim to optimize solar and battery energy use across all base stations. Therefore, our approach is expected to outperform YARN in energy management and resource utilization.

We also modified several existing approaches that address similar problems and compared their performance with our solution. These modified approaches are detailed below.

\textbf{(a) Slack Aware Non-Preemptive EDF (NPEDF):}
        In this approach by Lee et al. \cite{sota1}, tasks are selected upon arrival based on their slack time, which refers to the time they can be delayed without missing the deadline.
This approach prioritizes tasks with minimal slack time. \textbf{(b) ASAP HUF:}
        Tasks with the highest utilization are given priority. \textbf{(c) ASAP LUF:}
        Tasks with the lowest utilization are given priority. \textbf{(d) Execute on Arrival (EA):}
        This approach handles tasks as they arrive, regardless of their deadlines. Either the task is executed upon arrival or dropped altogether. The approach prioritizes tasks based on maximum revenue. We implemented this approach from Chouikhi et al. \cite{sota2}.

\subsection{Comparisons of Proposed Approach for different battery capacity}
\begin{figure}[tb!]
    \centering
    \includegraphics[scale = .55]{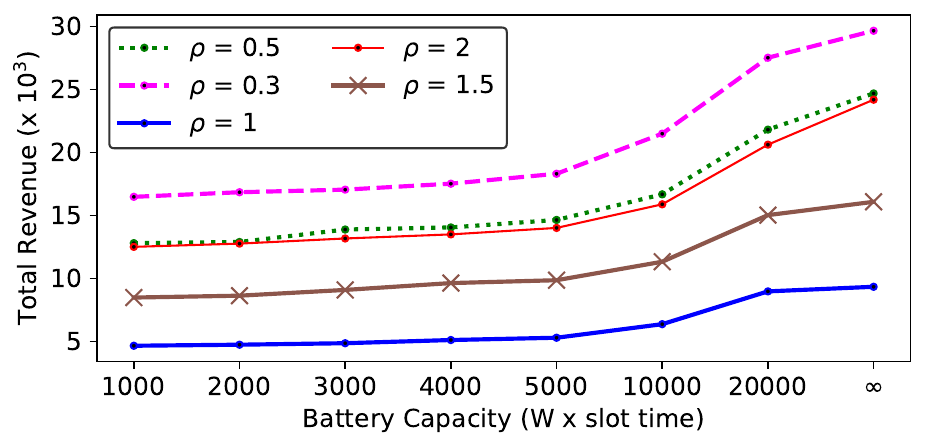}
    \caption{\footnotesize Variation of total revenue against battery capacities for different values of $\rho$ in real-life dataset \cite{tang,solar_power_generation_data}}
    \label{result:battery}
\end{figure}

\autoref{result:battery} shows the variation of the VES's total revenue varies when the battery's capacity changes. As the capacity of the battery increases, the total revenue increases and becomes maximum when the battery capacity is unlimited. The result is expected since if the battery capacity increases, the amount of battery storage for future usage increases, which makes the execution of the task power efficient, increasing the revenue of the VES.

\subsection{Comparison of Offline Approach for different values of $\rho$}

 We explored the relationship between the ratio $\rho$ (average utilization $u_i$ divided by the average deadline ($d_i$ - $a_i$)) and total revenue. It is important to note that our revenue is directly influenced by the average utilization and inversely affected by the average deadline.

\begin{figure}[tb!]
    \centering
    \includegraphics[scale = .55]{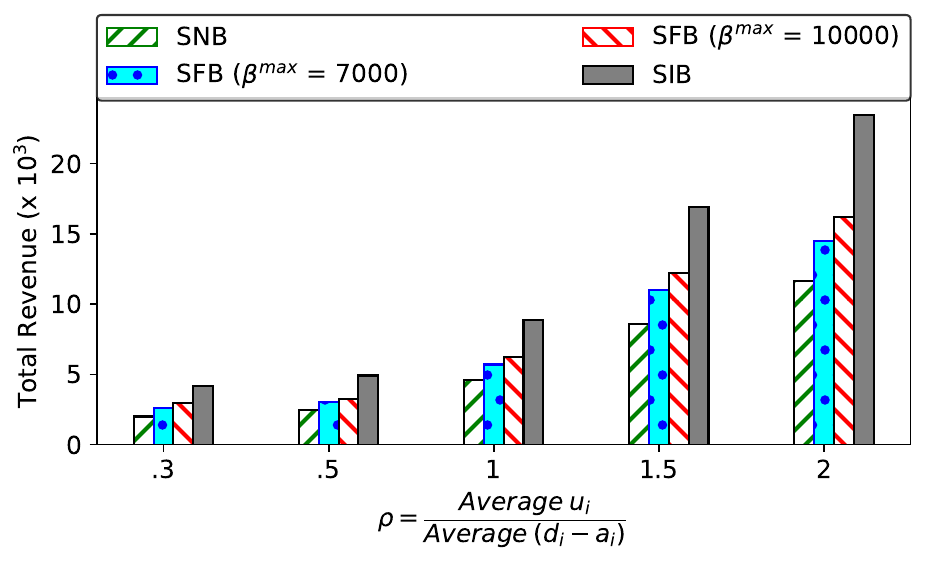}
    \caption{Variation of total revenue against different values of $\rho$ for different offline approaches in a single day time frame in real-life dataset \cite{tang,solar_power_generation_data}}
    \label{fig:offline}
\end{figure}

The comparison, shown in \autoref{fig:offline}, was experimented with across different $\rho$ values to analyze the total revenue. We considered various offline approaches, including scenarios with no battery, limited battery capacities of 10000 and 7000, and unlimited battery capacity. We observed a consistent trend across all values of $\rho$: as battery capacity increased, so did total revenue. This trend is significant as it underscores the role of energy storage in enhancing profitability.

\begin{figure}[tb!]
    \centering
    \includegraphics[scale = .55]{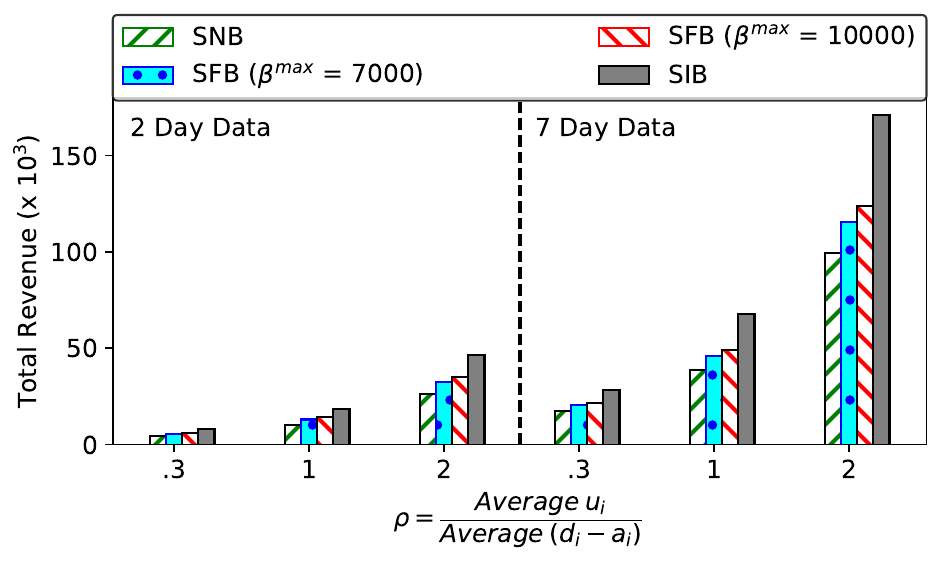}
    \caption{Variation of total revenue against different values of $\rho$ for different offline approaches in a multiple days time frame in real-life dataset \cite{tang,solar_power_generation_data}}
    \label{fig:multiple_day}
\end{figure}

A notable finding from the graph is that as $\rho$ values increase, the total revenue of all offline scenarios also increases. This phenomenon can be attributed to two key factors. Firstly, tasks with shorter deadlines tend to yield more revenue on average, creating opportunities for increased revenue. Secondly, tasks with higher utilization rates contribute significantly to revenue generation. As a result, higher $\rho$ values lead to a greater overall revenue across different offline approaches. Similarly, in the case of multiple days, as in \autoref{fig:multiple_day}, the total revenue increases as $\rho$ increases.

\subsection{Comparison of online approaches for different ranges of variation from predicted value} \label{subsection}
\begin{figure}[tb!]
    \centering
    \includegraphics[scale=.55]{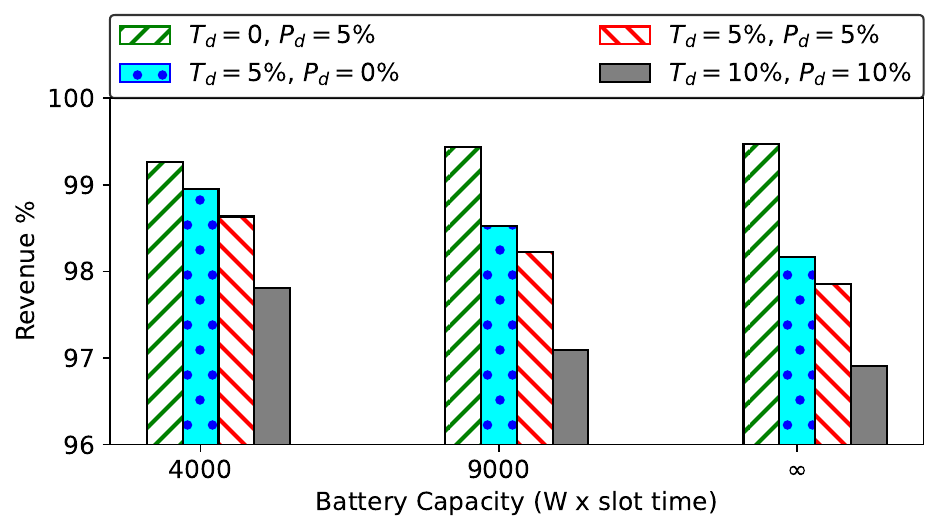}
    \caption{Variation of revenue\% against different battery capacities for online approaches with multiple ranges of variation from predicted value in real-life dataset \cite{tang,solar_power_generation_data}}
    \label{fig:deviation}
\end{figure}

\autoref{fig:deviation} shows the revenue\% of VES as the variation in real-time task increases from the predicted task trace and power profile. The revenue is presented in terms of revenue\%. We calculated the revenue\% considering the revenue from the offline approach as 100\%. We compared revenue percentages across different task deviations $T_d$ from the predicted task set $I$ and various power deviations $P_d$ from the predicted power $S_t^{actual}$. We explored this analysis across different battery capacities to understand how the revenue percentage varies between the online and offline approaches. 

As $T_d$ increases, the revenue percentage decreases accordingly, indicating greater task deviation. Notably, at $T_d$ and $P_d$ close to 10\%, the revenue percentage reaches 98\%. Despite variations in task and solar power, the difference in revenue percentage remains within 3\%, showcasing the robust performance of our algorithm in the online scenario. The deviation is minimal because, although the revenue may decrease when the task or power levels fall below predicted values, the system can also accommodate new, unallocated tasks that arise. Our approach is robust enough to integrate these unforeseen tasks, significantly contributing to overall profitability.

\subsection{Comparisons of Proposed Approach with State-of-Art-Approaches}
 
\begin{figure}[tb!]
    \centering
    \includegraphics[scale = .55]{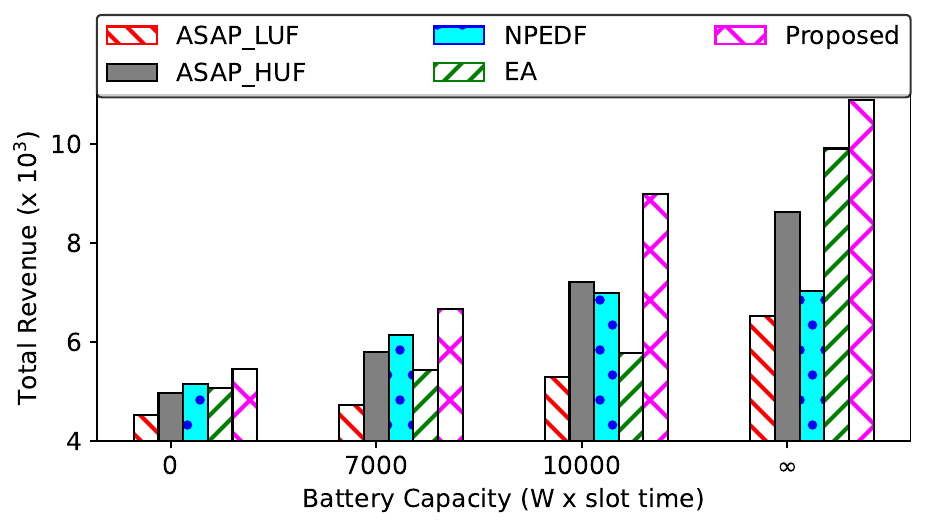}
    \caption{Variation of total revenue against different battery capacity for different state-of-art approaches for single day time frame in real-life dataset \cite{tang,solar_power_generation_data}}
    \label{fig:sota_1}
\end{figure}

We comprehensively compared our approach against several state-of-the-art methods, both with and without batteries, across varying battery capacities, as illustrated in \autoref{fig:sota_1}.

One key observation is that our method performs admirably even in scenarios without batteries. However, the true strength of our algorithm lies in its integration with battery technology, showcasing a substantial increase in profitability, as anticipated. Our approach strategically delays task execution using static information and employs power-shifting techniques, which are pivotal for its enhanced performance.

\begin{figure}[tb!]
    \centering
    \includegraphics[scale = .55]{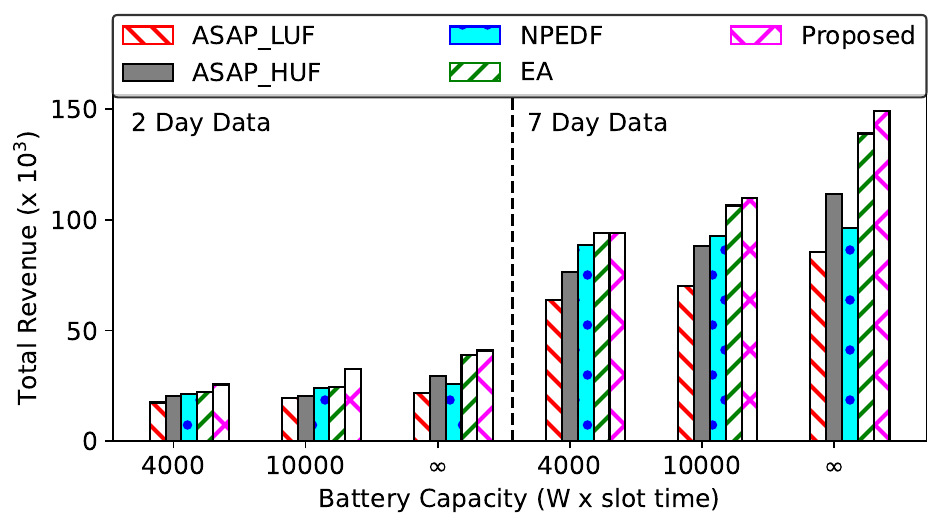}
    \caption{Variation of total revenue against different battery capacity for different state-of-art approaches for multiple days time frame in real-life dataset \cite{tang,solar_power_generation_data}}
    \label{fig:sota_multipleday}
\end{figure}

As shown in \autoref{fig:sota_1}, our approach exhibits improved efficacy with increasing battery capacities, highlighting its scalability and adaptability to varying energy storage constraints. Similarly, we have compared our approach on multiple-day scenarios as shown in \autoref{fig:sota_multipleday}, and our approach does well even in this scenario.

\subsection{Comparison between online and offline approaches}
\begin{figure}[tb!]
    \centering
    \includegraphics[scale = .55]{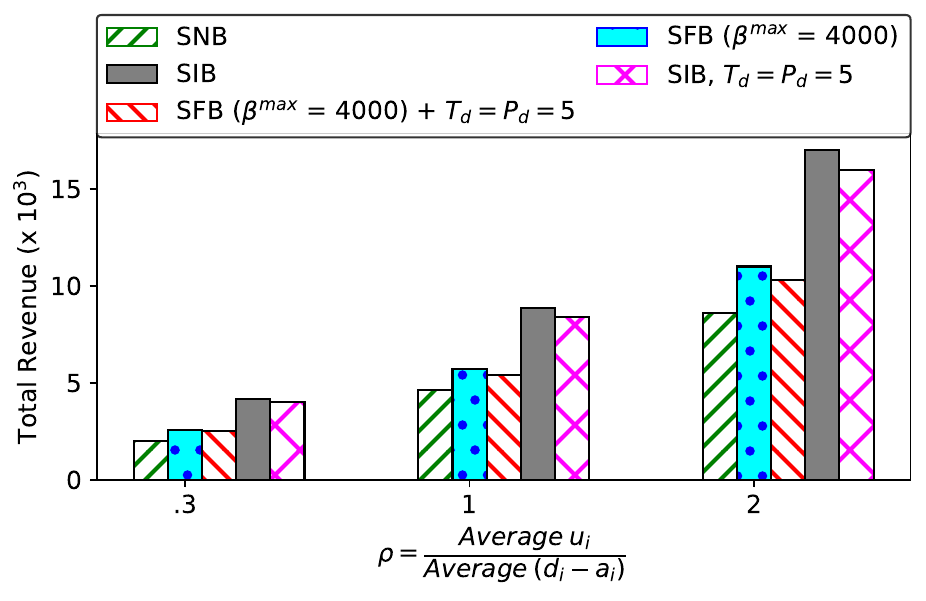}
    \caption{Variation of total revenue against different values of $\rho$ for different online and offline approaches in real-life dataset \cite{tang,solar_power_generation_data}}
    \label{fig:online_offline}
\end{figure}
As shown in \autoref{fig:online_offline}, a notable observation is the decrease in revenue for VES when transitioning from an offline to an online mode of execution. This decline is intuitive since the offline mode of our approach benefits from a comprehensive analysis of the global task pool and power profile before making decisions, a privilege not afforded to the online scenario. The results showcase minimal disparity between online and offline scenarios as explained previously in \autoref{subsection}. Also, as $\rho$ increases, there's a noticeable uptick in total revenue for both execution modes, aligning with our previous findings.

\subsection{Handling Robustness}

\begin{figure}
    \centering
    \includegraphics[scale=0.5]{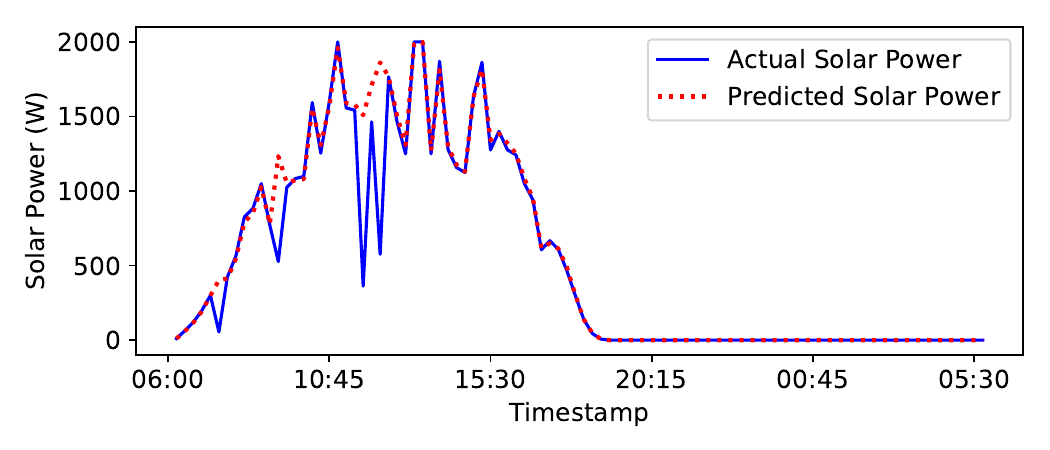}
    \caption{Spikes in incoming solar data}
    \label{fig:powerspike}
\end{figure}

\begin{figure}
    \centering
    \includegraphics[scale=0.5]{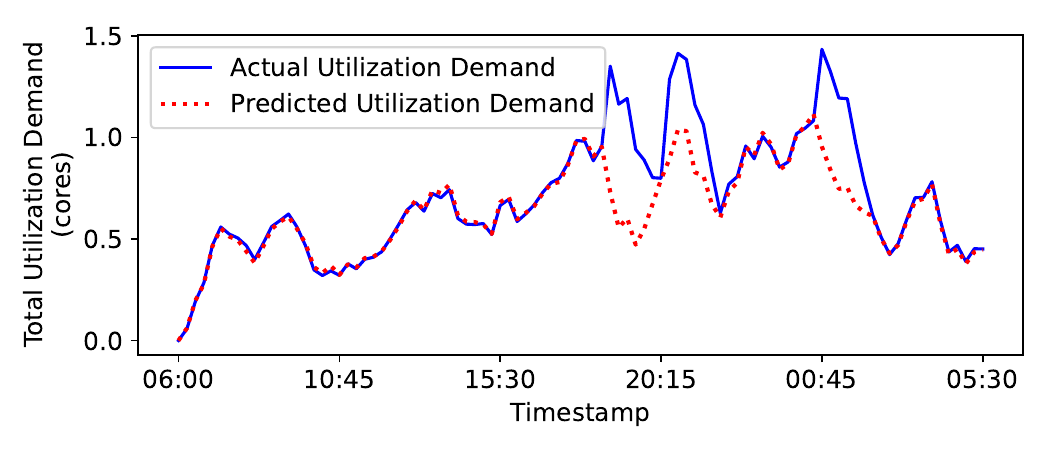}
    \caption{Spikes in coming task data}
    \label{fig:taskspikes}
\end{figure}

    Our proposed approach demonstrates robustness. To illustrate this, we intentionally introduce spikes into the incoming solar power data and task data. For the solar power data, we randomly select time slots and impose sudden drops in power from the predicted values. Similarly, we choose specific time slots for the incoming task data and add a substantial number of tasks compared to the predicted data. The spikes for the solar power data and incoming task data are shown in \autoref{fig:powerspike} and \autoref{fig:taskspikes}, respectively.

\begin{figure}
    \centering
    \includegraphics[scale=0.5]{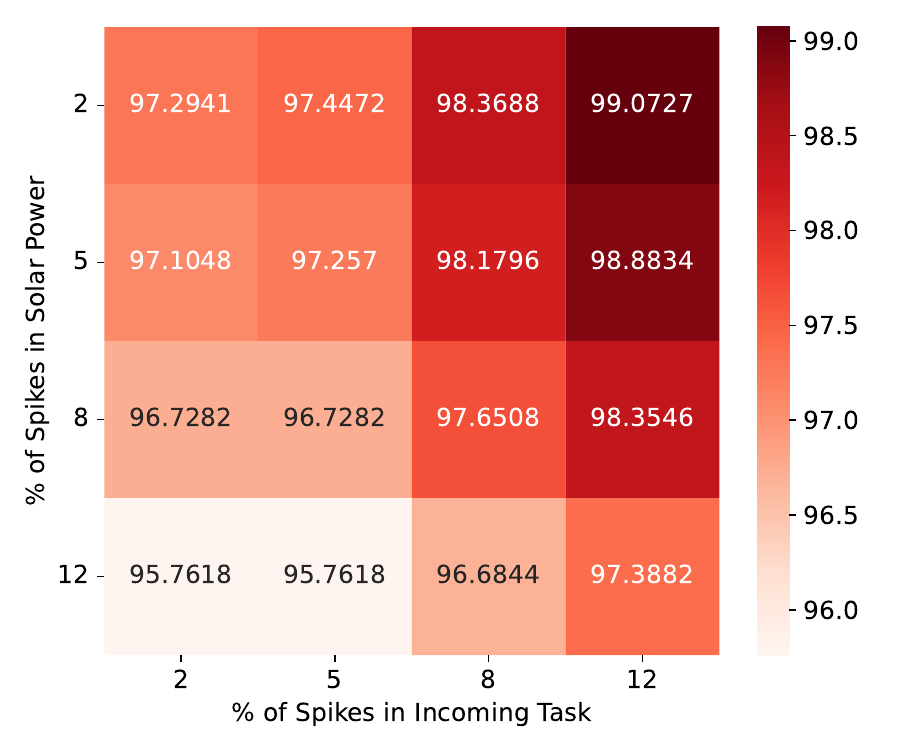}
    \caption{Variation of revenue \% of the proposed approach in different quantity of spikes in incoming solar and task data}
    \label{fig:robustness}
\end{figure}

In \autoref{fig:robustness}, we present the robustness of our approach. We calculate the revenue percentage considering a baseline scenario with zero spikes and a 5\% deviation from the predicted data. The decrease in percentage values reflects the reduction relative to the zero-spike scenario. To evaluate the performance of our approach, we apply a 5\% deviation from the predicted data and introduce spikes of 2\%, 5\%, 8\%, and 12\% to both the incoming task and solar power time slots, which the ES has no indication previously, and it deals with the situation instantaneously.

As the number of spikes in the solar power data increases, the overall revenue percentage decreases, as the total incoming power is reduced due to these spikes. Consequently, with lower overall incoming solar power, the total revenue diminishes, leading to a decrease in revenue percentage. Conversely, when we increase the number of spikes in the incoming tasks, the total revenue rises, as a larger pool of tasks allows the energy system to make better selections, resulting in greater revenue. Thus, the total revenue increases, leading to an increase in revenue percentage. Notably, even with a significant number of spikes, the revenue percentage decreases by a maximum of only 4\%, indicating that our approach is indeed robust.

\section{Real-life Application}

Solar energy alone is not entirely reliable due to its inconsistency, which is why it is supplemented with energy from the electric grid. For cloud providers, this approach is advantageous as it reduces the overall electricity consumption from the grid. However, relying solely on solar energy makes the cloud provider less dependable, as tasks cannot be completed when depleted solar energy. Since solar energy is limited, there are often more tasks than available energy.

To address this, we consider a setup where the cloud provider operates multiple edge servers: some powered exclusively by the grid and others by solar energy. The overall energy cost can be reduced by prioritizing the use of solar energy for most tasks and using grid energy for the remainder. This strategy provides a practical solution for real-world implementation, balancing energy efficiency and task reliability.
\section{Conclusion}
Our research underscores strategic energy management's significance in enhancing profitability and sustainability within Vehicular Edge Systems (VES). By introducing an energy-aware scheduling approach that integrates solar energy utilization with battery storage and adaptive task execution, we have demonstrated considerable advancements in optimizing task scheduling and power allocation. Our work maximizes revenue by leveraging knowledge of task sets and solar energy profiles across different timeslots and showcases adaptability in handling online scenarios with slight deviations from predicted data.

Incorporating task migration features to achieve better load balancing across edge servers could be a reasonable extension of this work. Additionally, it includes grid energy sources, thus offering a more comprehensive and versatile energy management solution.

\newpage

\setstretch{0.85}
\footnotesize
\bibliographystyle{IEEEtran}
\bibliography{mtp}
\end{document}